\documentclass{article} 
\usepackage{iclr2027_conference,times}

\usepackage{hyperref}
\usepackage{url}

\usepackage{amsmath,amssymb}
\usepackage{physics}
\usepackage{graphicx}
\usepackage{booktabs}
\usepackage{multirow}
\usepackage{xcolor}
\usepackage{tikz}
\usepackage{bm}
\usepackage{cleveref}

\crefname{equation}{Eq.}{Eqs.}
\Crefname{equation}{Eq.}{Eqs.}

\newcommand{\D}{\mathcal{D}}
\newcommand{\Dt}{\widetilde{\mathcal{D}}}
\newcommand{\loss}{\mathcal{L}}
\newcommand{\Cload}{C_{\mathrm{load}}}
\newcommand{\sd}[1]{{\scriptsize$\pm$#1}}    

\title{Distilling Datasets into Shallow Circuits\\ for Quantum Machine Learning}

\author{Guang Lin, Qibin Zhao  \\
RIKEN AIP, guang.lin@riken.jp
}

\iclrfinalcopy 
\begin{document}

\maketitle

\begin{abstract}
In quantum machine learning, training a quantum model requires each sample to be prepared as a quantum state by a loading circuit that must be re-executed for every shot at every training step. The total burden therefore scales with both the number of samples and the cost of preparation. Existing approaches reduce the loading cost of individual inputs or distill data and compress their representations before applying a separate quantum encoding. This separation can leave resulting samples costly to prepare or cause additional loss of the information during subsequent compilation into shallow circuits. We propose quantum dataset distillation (QDD), which distills the full dataset directly into a small set of shallow circuits. Each synthetic sample is parameterized as a staircase circuit corresponding to a low-rank tensor network, making the sample and its loading circuit the same object. The circuit parameters are optimized classically with exact gradients using distribution matching under an explicit loading budget, without post-hoc state-preparation synthesis. On MNIST and Fashion-MNIST, only 10 circuits per class ($0.0017\times$ the full dataset size) achieve accuracy comparable to full-data training and outperform selection baselines under matched sample and loading budgets. In finite-shot training, QDD reaches 95\% of the full-data accuracy with more than $100\times$ fewer cumulative shots. Additionally, we validate QDD on real quantum hardware, demonstrating its practical deployment potential.
\end{abstract}

\section{Introduction}
\label{sec:intro}

Quantum machine learning (QML) uses quantum circuits to represent and optimize learning models, providing a general framework for developing quantum-enhanced learning algorithms \citep{schuld2015introduction,biamonte2017quantum}. A growing body of theoretical and empirical studies has explored the potential of QML, suggesting that quantum models can capture meaningful structure in classical data \citep{mitarai2018quantum,caro2022generalization,devadas2025quantum}.

Despite this progress, training quantum models on realistically sized datasets remains prohibitively expensive, and a central obstacle lies not in the model itself, but in the data interface. Before a quantum model can process a classical sample, a loading circuit must prepare its corresponding quantum state. Exact amplitude encoding of an arbitrary input generally requires exponentially many two-qubit gates in the number of qubits \citep{mottonen2004transformation,plesch2011quantum,schuld2018supervised}. This preparation must be repeated for every measurement shot whenever the sample is used during training, so the resulting cost therefore grows with both the number of training samples and the cost of preparing each one. For datasets such as MNIST, which contains sixty thousand training images, repeated loading creates a substantial quantum execution burden, motivating training sets that are both compact and inexpensive to prepare.

In response, several approaches have been developed to alleviate this bottleneck. Approximate amplitude encoding and tensor-network compilation reduce the preparation cost of individual inputs by compressing each sample independently, typically under a reconstruction objective \citep{nakaji2022approximate,dilip2022data}. In parallel, classical dataset distillation synthesizes compact training sets from full datasets \citep{wang2018dataset,yu2023dataset}, offering a natural way to reduce the number of samples loaded into a quantum model. A recent study combined these ideas by distilling representations within the classical component of a hybrid model and then applying a separate quantum encoding \citep{phalak2025dataset}. However, treating dataset construction and state preparation separately leaves the distilled representations unconstrained by the circuits used to prepare them. Even a distilled set may therefore remain costly to load or lose additional information during subsequent compilation into shallow circuits. The key challenge is to distill information from the full dataset directly into states that the target circuit family can prepare, preserving downstream utility under explicit sample and loading budgets.

\begin{figure}[t]
\begin{center}
\includegraphics[width=\linewidth]{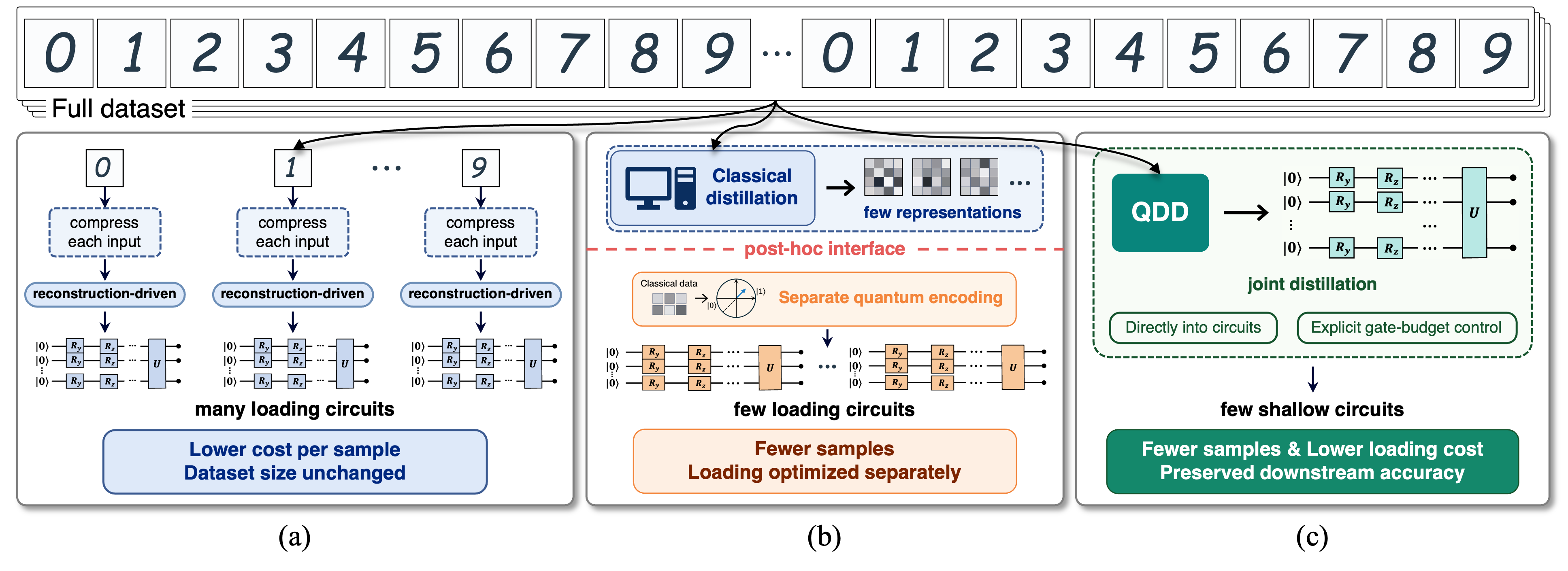}
\end{center}
\vskip -12pt
\caption{Three approaches to quantum training data. (a) Per-sample preparation. (b) Hybrid distillation with a separate encoding. (c) Quantum dataset distillation (QDD).}
\label{fig:overview}
\end{figure}

We address this gap with quantum dataset distillation (QDD), which directly distills the information from a full classical dataset into a compact training set of shallow circuits (\Cref{fig:overview}). QDD parameterizes each synthetic sample as a staircase circuit realizing a low-rank quantized tensor-train (QTT) state, so the sample and its circuit are the same object. This construction aligns the representation optimized during distillation with the one used for downstream training, avoiding a separate state-preparation approximation. The circuit parameters are optimized classically with exact gradients by matching class-conditional mean features of the downstream quantum model and mean density matrices. The staircase depth enforces an explicit gate budget, allowing QDD to preserve training utility within the available circuit capacity. After distillation, the resulting circuits form a reusable training set: their parameters remain fixed while only the downstream quantum model is trained.

We evaluate QDD on MNIST and Fashion-MNIST, comparing its accuracy with six baselines under matched sample and loading budgets and examining the benefits of joint optimization over distill-then-compress across gate budgets. We further measure cumulative shot costs during finite-shot training and validate hardware inference on three real quantum hardware.
In summary, our contributions are as follows:
\begin{itemize}
\item We introduce QDD, a framework that jointly addresses dataset size and quantum preparation cost by distilling training data directly into executable loading circuits, without post-hoc state-preparation synthesis.
\item We develop a staircase circuit parameterization of QTT states and optimize it classically with exact gradients, matching features and mean input states under an explicit gate budget.
\item We demonstrate that QDD retains strong performance with compact training sets and reduces cumulative shots in finite-shot training. Experiments on three IBM quantum processors further validate the deployment of QDD-trained models through shallow test loaders.
\end{itemize}

\section{Related work}
\label{sec:related}

\textbf{Training cost and data reduction in quantum machine learning (QML).}
Variational QML relies on repeated circuit executions to estimate losses and gradients \citep{mitarai2018quantum,schuld2019evaluating}, making training cost sensitive to both the number of samples processed and their state-preparation cost. Generalization analyses characterize the sample requirements of quantum models \citep{caro2022generalization}, but do not directly construct compact training sets from a given dataset. Quantum coreset methods address dataset reduction by selecting weighted subsets that approximate the original learning problem \citep{huang2024coreset}. Such selection reduces the number of retained samples while leaving their individual preparation costs unchanged unless a separate compression step is applied. QDD instead learns synthetic samples directly as shallow loading circuits, preserving the information while controlling both training-set size and preparation cost.

\textbf{Efficient quantum data loading.}
Exact preparation of arbitrary amplitude-encoded states generally requires exponentially many two-qubit gates in the number of qubits \citep{mottonen2004transformation,plesch2011quantum,schuld2018supervised}. Approximate amplitude encoding reduces this cost by fitting shallow variational circuits to individual target states \citep{nakaji2022approximate}. Tensor-network approaches exploit low-rank structure, representing inputs as matrix product states or quantized tensor trains and mapping these representations to sequential quantum circuits \citep{ran2020encoding,dilip2022data,lin2026structured}. These methods primarily seek faithful, inexpensive preparation of given inputs, leaving the training-set size unchanged. QDD builds on structured staircase circuits to synthesize a compact training set, optimizing the information retained across samples for downstream learning under a loading budget.

\textbf{Dataset distillation for QML.}
Classical dataset distillation synthesizes compact training sets that preserve the learning utility of larger datasets \citep{wang2018dataset,yu2023dataset}. Synthetic samples are typically optimized as images or feature representations, without accounting for quantum state-preparation cost. In QML, these samples must subsequently be encoded as quantum states. \citet{phalak2025dataset} explored dataset distillation for a hybrid quantum-classical model, combining distilled classical representations with a separate quantum encoding. This reduces the training-set size but does not explicitly constrain the quantum loading cost during distillation.

A natural extension is to compile the distilled samples into shallow preparation circuits, reducing both sample count and loading cost. We evaluate this strategy through a matched-objective distill-then-compress baseline (\Cref{sec:results-posthoc}). Its effectiveness depends on whether compilation preserves the information retained during distillation, which can be difficult under tight gate budgets. QDD incorporates the loading constraint from the outset: it optimizes synthetic samples directly as shallow circuits, preserving downstream utility under an explicit two-qubit gate budget.

\section{Background}
\label{sec:background}

\textbf{Quantum states and low-rank circuit representation.}
An $n$-qubit circuit acts on a state $\ket{\psi} \in (\mathbb{C}^{2})^{\otimes n}$ through a sequence of unitary gates; we use the computational basis $\{\ket{s} : s \in \{0,1\}^{n}\}$, $R_y$ rotations and CNOTs as native gates, and standard notation \citep{nielsen2000quantum}. Applying a circuit $U$ to $\ket{0}^{\otimes n}$ and measuring in the computational basis returns the bit string $s$ with probability
\begin{equation}
p(s) \;=\; \left|\bra{s} U \ket{0}^{\otimes n}\right|^{2} \;=\; \bra{s}\rho\ket{s},
\qquad
\rho = \ketbra{\psi},
\qquad
\ket{\psi} = U\ket{0}^{\otimes n},
\end{equation}
and the expectation of an observable $A$ is $\Tr[A\rho]$, which is linear in $\rho$. On quantum hardware, probabilities and expectation values are estimated from $S$ measurement shots, each of which re-executes the complete circuit, including the preparation of its input state. We measure circuit cost by the number of two-qubit gates $N_{2q}$; the circuit $E$ that prepares a sample is its loading circuit, with loading cost $\Cload(E) = N_{2q}(E)$, paid again at every shot.

Indexing the amplitude vector $\psi \in \mathbb{R}^{2^{n}}$ by the $n$ bits of $s$ turns it into an order-$n$ tensor. Its tensor-train decomposition is a matrix-product state (MPS), referred to as a quantized tensor train (QTT) in this binary representation \citep{oseledets2011tensor,khoromskij2011d}. Natural images often admit accurate low-rank approximations through TT-SVD truncation, making shallow staircase circuits a natural loading family, as developed in \Cref{sec:parameterization}.

\textbf{Training budget.}
The downstream quantum model is a parameterized circuit $W_\theta$ with $P$ trainable angles. On hardware its gradients are estimated by the parameter-shift rule \citep{mitarai2018quantum,schuld2019evaluating}, which evaluates $2P + 1$ circuits per sample, so $T$ steps on batches of $b$ samples at $S$ shots cost
\begin{equation}
  N^{\mathrm{train}}_{\mathrm{shots}} \;=\; (2P+1)\, b\, S\, T .
\label{eq:budget}
\end{equation}
Every execution begins with the loading circuit of its sample. The training set therefore determines the training cost through two quantities: the number of samples, and the loading cost at every shot.

\textbf{Dataset distillation.}
Dataset distillation replaces a training set $\D$ of $n_{\mathrm{data}}$ samples with a much smaller distilled set $\Dt$ that retains its training utility \citep{wang2018dataset,yu2023dataset}. For $C$ classes and $\mathrm{IPC}$ synthetic images per class, the distilled set contains $M = C \cdot \mathrm{IPC} \ll n_{\mathrm{data}}$ samples. Classically, presenting a synthetic sample to the model is free; in QML, every synthetic sample must be realized by a state-preparation circuit whose cost is paid at every shot.

\section{Method}
\label{sec:method}

Quantum dataset distillation (QDD) produces a small labeled set of executable loading circuits. It controls two resources explicitly: the number of distilled samples and the gate cost of preparation. We first formulate this joint objective, then describe the circuit family, the distillation loss, and the offline distillation and deployment procedure.

\subsection{Problem formulation}
\label{sec:problem}

Let $\D=\{(x_i,y_i)\}_{i=1}^{n_{\mathrm{data}}}$ be a labeled dataset, and let $\rho_i$ denote the density operator of the quantum state representing $x_i$ through the low-rank interface. QDD constructs
\begin{equation}
    \Dt(\Phi,\widetilde Y)
    =
    \bigl\{\bigl(E(\phi_j),\tilde y_j\bigr)\bigr\}_{j=1}^{M},
    \qquad
    \tilde\rho_j
    =
    E(\phi_j)\ketbra{0^{\otimes n}}E(\phi_j)^\dagger,
    \qquad
    M\ll n_{\mathrm{data}},
\end{equation}
where $\Phi=\{\phi_j\}_{j=1}^{M}$ contains the loading-circuit parameters and $\widetilde Y=\{\tilde y_j\}_{j=1}^{M}$ the class labels used for downstream training. The mean preparation cost is
\begin{equation}
    \overline{\Cload}(\Phi)
    =
    \frac{1}{M}\sum_{j=1}^{M}N_{2q}\!\left(E(\phi_j)\right).
\end{equation}
For a sample budget $M$ and an average loading budget $B_{2q}$, the loading-aware distillation problem is
\begin{equation}
    \min_{\Phi}\ \loss_{\mathrm{distill}}(\Phi;\D)
    \qquad
    \text{subject to}
    \qquad
    \overline{\Cload}(\Phi)\le B_{2q}.
\label{eq:qdd_problem}
\end{equation}
The two budgets act on these two quantities: $M$ bounds how many samples a full-batch training step must visit, and $B_{2q}$ bounds the average gate cost of preparing each of them. Crucially, no projection or state-preparation synthesis is applied after distillation.

\subsection{Circuit-native synthetic samples}
\label{sec:parameterization}

Each synthetic sample is parameterized as an $L$-layer sequential circuit, in which nearest-neighbor two-qubit blocks are arranged in a staircase pattern. Define an initial rotation layer $R(\phi_j^{(0)})$ and the $l$-th staircase layer $S(\phi_{j,l})$ by
\begin{equation}
\begin{aligned}
    R(\phi_j^{(0)})=\bigotimes_{q=0}^{n-1}R_y\!\left(\phi_{j,q}^{(0)}\right),\quad &S(\phi_{j,l})=B(\phi_{j,l,1})_{1,0}B(\phi_{j,l,2})_{2,1} \cdots B(\phi_{j,l,n-1})_{n-1,n-2},\\
    &E(\phi_j)=S(\phi_{j,L})\cdots S(\phi_{j,1})R(\phi_j^{(0)}).
\end{aligned}
\label{eq:staircase}
\end{equation}
The rightmost operation acts first, so each $S(\phi_{j,l})$ sweeps from bond $(n-1,n-2)$ to bond $(1,0)$. Every two-qubit block contains six trainable angles and two CNOTs:
\begin{equation}
    B(a,\ldots,f)=
    \bigl[R_y(e)\otimes R_y(f)\bigr]\,\mathrm{CNOT}\,
    \bigl[R_y(c)\otimes R_y(d)\bigr]\,\mathrm{CNOT}\,
    \bigl[R_y(a)\otimes R_y(b)\bigr].
\label{eq:block}
\end{equation}
At zero angles, the two CNOTs cancel and $B$ is the identity. A full $L$-layer loader therefore has the exact cost
\begin{equation}
    \Cload(\phi_j)=2(n-1)L.
\label{eq:loadcost}
\end{equation}
The circuit can be embedded along a connected nearest-neighbor chain without SWAP gates. Because the target amplitudes are real, the loaders use only $R_y$ rotations and CNOTs. This staircase construction is closely related to low-rank MPS preparation: an $L$-layer circuit produces an MPS with bond dimension at most $2^L$. Thus, circuit depth controls both representational capacity and loading cost. More importantly, the optimized sample is already a gate-level loading circuit rather than a state that must be compiled afterward (\Cref{app:compilation}).

\subsection{Distillation objective}
\label{sec:objectives}

QDD combines distribution matching in the representation of the quantum model with direct matching in state space. Both terms can be optimized without training the downstream model.

\textbf{Distribution matching (DM).}
For a quantum model $W_\theta$, define the quantum feature map
\begin{equation}
    f_\theta(\rho)=
    \left(\bra{s}W_\theta\rho W_\theta^\dagger\ket{s}\right)_{s\in\{0,1\}^n},
\end{equation}
which collects the computational-basis measurement distribution after $W_\theta$ is applied to $\rho$. For class $c$, the mean features of the real and synthetic data are
\begin{equation}
    \mu_c^{\mathrm{real}}(\theta)=
    \frac{1}{|\D_c|}\sum_{i\in\D_c}f_\theta(\rho_i),
    \qquad \mu_c^{\mathrm{syn}}(\theta)=
    \frac{1}{|\Dt_c|}\sum_{j\in\Dt_c}f_\theta(\tilde\rho_j).
\end{equation}
We match these class-conditional means across randomly initialized models:
\begin{equation}
    \loss_{\mathrm{DM}}=
    \mathbb{E}_{\theta\sim p(\theta)}\sum_{c=1}^{C}
    \left\lVert \mu_c^{\mathrm{real}}(\theta)-\mu_c^{\mathrm{syn}}(\theta) \right\rVert_2^2,
\label{eq:dm}
\end{equation}
where $p(\theta)$ is a Gaussian prior over the circuit angles. The sampled circuits are not trained: their parameters are resampled during distillation, and gradients are taken only with respect to the synthetic-circuit parameters $\Phi$. Consequently, $\loss_{\mathrm{DM}}$ emphasizes differences between real and synthetic states that are visible to the target model family.

\textbf{Density-matrix matching.}
Each component of $f_\theta$ can be written as
$\Tr[W_\theta^\dagger\ketbra{s}W_\theta\rho]$
and is therefore linear in $\rho$. Hence, the class-conditional mean feature depends only on the mean input state, given by
\begin{equation}
    \bar\rho_c^{\mathrm{real}}=\frac{1}{|\D_c|}\sum_{i\in\D_c}\rho_i,
    \qquad
    \bar\rho_c^{\mathrm{syn}}=\frac{1}{|\Dt_c|}\sum_{j\in\Dt_c}\tilde\rho_j.
\end{equation}
We directly match these class-mean states:
\begin{equation}
    \loss_\rho=\sum_{c=1}^{C}
    \left\lVert \bar\rho_c^{\mathrm{real}}-\bar\rho_c^{\mathrm{syn}} \right\rVert_F^2.
\label{eq:rho}
\end{equation}
The small distilled set and restricted loading-circuit family generally prevent the two mean states from matching exactly. In this constrained regime, $\loss_\rho$ provides a model-independent measure of their discrepancy in the full state space, whereas $\loss_{\mathrm{DM}}$ emphasizes the measurement projections exposed by the downstream architecture. Their complementary effects are evaluated in \Cref{app:objectives}.
The resulting distillation objective is
\begin{equation}
    \loss_{\mathrm{QDD}}(\Phi)=
    \loss_{\mathrm{DM}}(\Phi)+\alpha\loss_\rho(\Phi).
\label{eq:objective}
\end{equation}
Together, the two terms encourage the synthetic circuits to reproduce both the features used by quantum models and the class-wise mean states of the real data. Since the circuits are optimized at the target depth, this information is distilled directly under the prescribed loading budget.

\subsection{Offline distillation and deployment}
\label{sec:training}

Each distillation step draws a batch of real examples and a random initialization $\theta$ of the quantum model $W_\theta$. We then evaluate \Cref{eq:objective} and update only the synthetic-circuit parameters $\Phi$; $W_\theta$ serves as a feature extractor and is resampled rather than trained. At the studied scale, all states and circuits are contracted exactly, and reverse-mode automatic differentiation computes exact gradients of the sampled objective. This offline stage requires no measurement shots and runs classically.

After distillation, each optimized loader $E(\phi_j)$ is exported directly as the circuit defined in \Cref{eq:staircase}. Because each synthetic sample is already an executable circuit, no classical reconstruction or separate state-preparation synthesis is required. Backend transpilation merely maps the exported circuit to the native gate set and connectivity of the target device.
During downstream training, every loading circuit $E(\phi_j)$ remains frozen, and only the parameters $\theta$ are updated. In our experiments, $\theta$ is trained in classical simulation, either with exact gradients or with parameter-shift gradients estimated from sampled finite shots \citep{schuld2019evaluating}.

\section{Experiments}
\label{sec:experiments}

We evaluate QDD along three main dimensions: its advantage over data selection under matched sample and loading budgets, the benefit of joint optimization over distill-then-compress across gate budgets, and its effectiveness in finite-shot parameter-shift training after offline distillation with exact gradients. We further report ablations and a hardware evaluation that follows the exported circuits through finite shots and device noise models to execution on three IBM quantum processors, with additional results in \Cref{app:extra}.

\subsection{Experimental setup}
\label{sec:setup}

\textbf{Tasks and interface.}
We study 10-class classification on MNIST and Fashion-MNIST. Images are mapped to normalized 8-qubit states in QTT order, corresponding to a resolution of $16\times16$. Real training and test images enter through a rank-8 interface: the truncated states are used directly in exact simulation and compiled into an $L=3$ staircase circuit for the noise and hardware evaluations. The same inference interface is used across all training sets. Preprocessing and compilation details are provided in \Cref{app:hyper,app:compilation}, and a $10$-qubit study appears in \Cref{app:n10}.

\textbf{Quantum model and distillation.}
All methods use the same quantum model $W_\theta$, consisting of $L_c=16$ brickwork sublayers with 112 CNOTs. QDD distills $M \in \{10,100,500\}$ circuits, corresponding to $\mathrm{IPC} \in \{1,10,50\}$. Unless otherwise stated, each distilled sample uses an $L=3$ staircase loader with 42 CNOTs. More details are given in \Cref{app:costmodel}.

\textbf{Baselines.}
We compare QDD with random, kernel herding \citep{chen2012super}, forgetting-based \citep{toneva2018empirical}, coreset \citep{huang2024coreset}, k-center greedy \citep{sener2017active}, and k-means selections; each selected image is compiled into the same staircase circuit. We additionally consider a distill-then-compress control, which optimizes unconstrained states with the QDD objective and subsequently compiles them to the same budget, isolating the effect of joint optimization. Full-data training provides the reference performance.

\textbf{Evaluation protocols.}
To assess predictive utility, we train a fresh quantum model on each training set using exact gradients and evaluate it on the full $10$k-image test set. Unless stated otherwise, the results are averaged over three training seeds. To assess training efficiency, we train models from scratch using finite-shot parameter-shift gradients at $S \in \{100,1000,10000\}$ shots per circuit, give every compact training set the same total shot budget, and record the cumulative shots required to reach a fixed fraction of the full-data accuracy. Exact simulation experiments support the accuracy comparisons, whereas finite-shot experiments support the training-cost comparisons. Optimizer and cost settings are provided in \Cref{app:hyper,app:costmodel}.

\textbf{Noise and hardware evaluation.}
Exported circuits are first verified against the simulated states and then evaluated through exact simulation, finite-shot sampling, archived IBM device noise models, and execution on three IBM quantum processors (\texttt{ibm\_kawasaki}, \texttt{ibm\_fez}, and \texttt{ibm\_boston}). These evaluations use 4096 shots and no error mitigation. Device configurations, shot counts, and transpilation details are provided in \Cref{app:hardware}.

\subsection{Distillation improves quantum training}
\label{sec:results-main}

\begin{table}[ht]
\caption{Test accuracy (\%) by images per class (IPC) on MNIST and Fashion-MNIST. All compact training sets use 42-CNOT loaders. Mean $\pm$ standard deviation over 3 training seeds.}
\label{tab:main}
\setlength{\tabcolsep}{7.3pt}
\begin{center}
\begin{tabular}{l ccc ccc}
\toprule
 & \multicolumn{3}{c}{MNIST} & \multicolumn{3}{c}{Fashion-MNIST} \\
\cmidrule(lr){2-4}\cmidrule(lr){5-7}
Method & IPC 1 & IPC 10 & IPC 50 & IPC 1 & IPC 10 & IPC 50 \\
\midrule
Random                      & 36.63\sd{2.04} & 68.18\sd{2.18} & 80.73\sd{0.73} & 52.65\sd{1.29} & 66.49\sd{1.11} & 69.81\sd{0.71} \\
\midrule
Herding                         & 57.18\sd{1.84} & 75.77\sd{2.58} & 79.96\sd{0.41} & 52.49\sd{0.10} & 67.23\sd{0.35} & 69.31\sd{1.00} \\
Forgetting & 9.21\sd{0.62} & 20.16\sd{1.09} & 24.71\sd{2.41} & 16.50\sd{0.83} & 19.03\sd{0.35} & 24.08\sd{0.85} \\
Coreset & 57.18\sd{1.84} & 77.62\sd{0.57} & 79.65\sd{0.26} & 52.49\sd{0.10} & 68.78\sd{0.17} & 68.81\sd{0.82} \\
K-center & 57.18\sd{1.84} & 36.98\sd{5.99} & 56.40\sd{2.81} & 52.49\sd{0.10} & 34.07\sd{2.40} & 48.05\sd{1.98} \\
K-means & 58.33\sd{3.29} & 79.67\sd{0.74} & 81.34\sd{1.01} & \textbf{58.44}\sd{0.53} & 68.92\sd{0.28} & 67.91\sd{0.91} \\
QDD (ours) & \textbf{64.04}\sd{1.07} & \textbf{81.62}\sd{0.33} & \textbf{82.25}\sd{0.37} & 58.06\sd{2.14} & \textbf{69.67}\sd{0.43} & \textbf{71.01}\sd{0.58} \\
\bottomrule
\end{tabular}
\end{center}
\vskip -9pt
\end{table}

\Cref{tab:main} reports the test accuracy of quantum models trained on compact training sets constructed by QDD and six baseline methods on MNIST and Fashion-MNIST. We compare these methods at IPC 1, 10, and 50, using the same downstream quantum model and a loading budget of 42 CNOTs per sample. QDD achieves the highest accuracy in five of the six settings; the exception is IPC 1 on Fashion-MNIST, where k-means leads by 0.38\%. With only 10 circuits per class, QDD reaches 81.62\% on MNIST and 69.67\% on Fashion-MNIST. Increasing to IPC 50 yields a slight improvement, suggesting that 100 distilled circuits already capture much of the information needed to train the chosen quantum model. The finer IPC sweep in \Cref{app:rtest} shows accuracy saturating between IPC 10 and 20, and the 10-qubit study in \Cref{app:n10} exhibits a similar pattern.

Unlike the baselines, QDD incorporates the loading constraint directly into the distillation process. Every baseline first chooses or averages real images and only afterward fits a 42-CNOT circuit to each of them, so the information that the circuit cannot represent is lost after the selection has been made. QDD optimizes the circuit parameters directly, so its samples are executable by construction and lose nothing at export. Additionally, QDD is not restricted to real images or their averages; its samples are jointly optimized to match their class-conditional mean features and mean density matrices, providing flexibility when only a few samples are available to represent each class. Accordingly, the gain over the strongest baseline is largest at IPC 1 on MNIST (5.71\%) and shrinks to about 1\% to 2\% at larger IPC, where several real samples already cover most of the class variability.

\subsection{Joint optimization vs. distill-then-compress}
\label{sec:results-posthoc}

\begin{table}[ht]
\vskip -5pt
\caption{Joint optimization vs. distill-then-compress on MNIST. Column headings give the loading budget in CNOTs per sample. Evaluation results are test accuracy (\%) over $3$ training seeds.}
\label{tab:posthoc}
\setlength{\tabcolsep}{7.2pt}
\renewcommand{\arraystretch}{0.95}
\begin{center}
\begin{tabular}{ll ccccc}
\toprule
IPC & Method & 14 & 28 & 42 & 56 & Avg. \\
\midrule
\multirow{3}{*}{1}  & Distill-then-compress & 26.70\sd{5.25} & 51.31\sd{1.78} & 60.99\sd{4.22} & 64.36\sd{1.18} & 50.84 \\
                    & Joint optimization & 29.44\sd{3.63} & 50.29\sd{1.85} & 64.04\sd{1.07} & 65.78\sd{1.70} & 52.39\\
                    & $\Delta$ & $\bm{+2.74}$ & $-1.02$ & $\bm{+3.05}$ & $\bm{+1.42}$ & $\bm{+1.55}$ \\
\midrule
\multirow{3}{*}{10} & Distill-then-compress & 59.32\sd{3.58} & 77.66\sd{1.08} & 80.57\sd{0.56} & 80.56\sd{0.42} & 74.53 \\
                    & Joint optimization & 68.57\sd{0.82} & 78.86\sd{0.49} & 81.62\sd{0.33} & 81.73\sd{0.71} & 77.70 \\
                    & $\Delta$ & $\bm{+9.25}$ & $\bm{+1.20}$ & $\bm{+1.05}$ & $\bm{+1.17}$ & $\bm{+3.17}$ \\
\midrule
\multirow{3}{*}{50} & Distill-then-compress  & 66.13\sd{1.44} & 81.40\sd{0.65} & 82.63\sd{0.45} & 82.43\sd{0.58} & 78.15 \\
                    & Joint optimization & 73.69\sd{0.84} & 80.12\sd{0.50} & 82.25\sd{0.37} & 83.08\sd{0.19} & 79.79 \\
                    & $\Delta$ & $\bm{+7.56}$ & $-1.28$ & $-0.38$ & $\bm{+0.65}$ & $\bm{+1.64}$ \\
\bottomrule
\end{tabular}
\end{center}
\end{table}

\Cref{tab:posthoc} compares joint optimization (QDD) with distill-then-compress on MNIST across IPC and loading budget. Both methods use the same distillation objective and downstream quantum model. They differ only in when the circuit constraint is imposed: QDD optimizes directly within the target circuit family, whereas distill-then-compress first optimizes unconstrained states and then approximates each of them with a circuit within the budget.

QDD provides its largest gains at the tightest loading budget. With 14 CNOTs per sample, it improves accuracy by 9.25\% at IPC 10 and 7.56\% at IPC 50. At IPC 10, QDD remains ahead at all four budgets, although the margin narrows to approximately 1\% from 28 CNOTs onward.
The narrowing gap is consistent with increasingly accurate post-hoc compilation. On MNIST at IPC 10, the mean overlap between the unconstrained distilled states and their compiled approximations rises from 0.76 at 14 CNOTs to 0.98 at 56 CNOTs (\Cref{app:compilation}). Under a tight budget, QDD can adapt the synthetic states to what the loading circuits can represent, whereas distill-then-compress commits to states that the circuits cannot reproduce and loses the difference at compilation.

\subsection{Training efficiency with finite shots}
\label{sec:results-econ}

\begin{figure}[ht]
\begin{center}
\includegraphics[width=\linewidth]{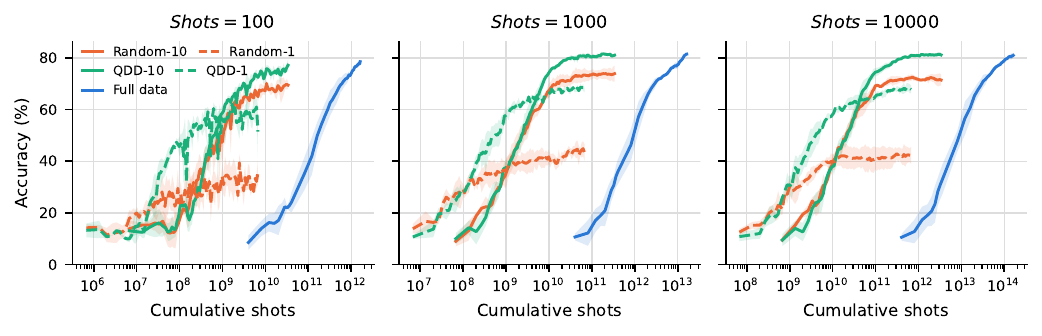}
\vskip -9pt
\caption{Finite-shot parameter-shift training on MNIST: test accuracy vs. cumulative shots.}
\label{fig:econ}
\end{center}
\end{figure}

\begin{table}[ht]
\vskip -12pt
\caption{Cumulative-shot savings (full-data/method; higher is better) at target fractions of full-data accuracy. n/a indicates targets not reached. The last row reports full-data shots in units of $10^{10}$.}
\label{tab:budget}
\setlength{\tabcolsep}{4.2pt}
\begin{center}
\begin{tabular}{l ccc ccc ccc}
\toprule
 & \multicolumn{3}{c}{Shots = 100} & \multicolumn{3}{c}{Shots = 1000} & \multicolumn{3}{c}{Shots = 10000} \\
\cmidrule(lr){2-4}\cmidrule(lr){5-7}\cmidrule(lr){8-10}
Training set & 60\% & 80\% & 95\% & 60\% & 80\% & 95\% & 60\% & 80\% & 95\% \\
\midrule
Random-1  & n/a & n/a & n/a & n/a & n/a & n/a & n/a & n/a & n/a \\
Random-10 & 207$\times$ & 185$\times$ & n/a & 529$\times$ & 238$\times$ & n/a & 667$\times$ & 259$\times$ & n/a \\
QDD-1     & $\bm{1905\times}$ & n/a & n/a & $\bm{2647\times}$ & 258$\times$ & n/a & $\bm{3333\times}$ & 205$\times$ & n/a \\
QDD-10 & 304$\times$ & $\bm{271\times}$ & $\bm{134\times}$ & 529$\times$ & $\bm{380\times}$ & $\bm{576\times}$ & 563$\times$ & $\bm{441\times}$ & $\bm{370\times}$ \\
\midrule
Full-data shots & 16 & 36 & 105 & 121 & 202 & 1050 & 1211 & 2019 & 7268 \\
\bottomrule
\end{tabular}
\end{center}
\vskip -3pt
\end{table}

\Cref{fig:econ} and \Cref{tab:budget} report finite-shot parameter-shift training on MNIST using circuits distilled with exact classical gradients. We train quantum models with $S\in\{100,1000,10000\}$ shots per circuit. Each training step processes the entire training set: 10 or 100 samples for the compact training sets and 60000 for the full-data reference. For each $S$, we measure the cumulative shots at which each averaged accuracy curve first reaches 60\%, 80\%, and 95\% of the best averaged full-data accuracy. QDD-10 is the only compact set tested that reaches the 95\% target at every shot setting, using 134$\times$, 576$\times$, and 370$\times$ fewer shots than full-data training at 100, 1000, and 10000 shots per circuit, respectively. Random-10 reaches the 80\% target but not the 95\% target, despite having the same sample count and per-step shot cost. With only ten samples, QDD-1 reaches the 60\% target at every setting and the 80\% target at 1000 and 10000 shots, whereas Random-1 never reaches the 60\% target within budget. At 1000 shots per circuit, QDD-10 also attains accuracy close to both full-data training (\Cref{tab:budget-minibatch}) and its exact-simulation reference (\Cref{tab:main}), showing that the reduction in measurement cost can be achieved while retaining strong predictive performance.

These results show that offline distillation preserves its training utility under finite-shot gradient estimation. The large savings relative to full-batch training arise partly from processing far fewer samples per step, while the comparison with random selection demonstrates the importance of the compact training set's training utility. A complementary comparison with full-data batch training at matched batch size is reported in \Cref{app:shot-results}. The distilled circuits can be reused across downstream training runs, with their one-time classical construction cost.

\subsection{Ablations}
\label{sec:results-ablations}

We examine sensitivity of QDD to multiple design choices. (a) Inference rank. We vary the QTT rank of inputs while keeping the distilled sets and trained quantum models fixed (\Cref{tab:interface-rank}). On MNIST at IPC 10, increasing the rank from 1 to 8 raises accuracy from 25.57\% to 81.62\%, with little further improvement at rank 16 (81.68\%). (b) Gate budget. Accuracy improves with the loading budget but largely saturates at 42 CNOTs per sample (\Cref{tab:posthoc}). On MNIST at IPC 10, increasing the budget to 56 CNOTs raises accuracy only 0.11\%, supporting $L=3$ as the default loading depth. (c) Distillation objective. Adding density-matrix matching to DM improves MNIST accuracy by 2.14\% at $L=2$ and 2.91\% at $L=3$ within IPC 1 (\Cref{tab:rho-ablation}), showing its value in this constrained setting. The effect is smaller at $L=3$ and IPC 10, as shown in \Cref{app:objectives}. Additional experiments examine sensitivity to numbers of shots, objective weight, and numbers of distilled samples (\Cref{app:extra}). These results motivate our default configuration for the main experiments.

\subsection{Noise and hardware evaluation}
\label{sec:hardware}
\begin{table}[htbp]
\vskip -12pt
\caption{Test accuracy (\%) on 128 MNIST images under exact simulation, ideal sampling, archived device noise models, and three IBM quantum processors. All sampled results use 4096 shots without error mitigation. The staircase rows compare training sets through identical 42-CNOT test loaders.}
\label{tab:hardware}
\setlength{\tabcolsep}{5pt}
\begin{center}
\begin{tabular}{ll cc ccc ccc}
\toprule
& & \multicolumn{2}{c}{Noiseless} & \multicolumn{3}{c}{Noise model} & \multicolumn{3}{c}{Real hardware} \\
\cmidrule(lr){3-4}\cmidrule(lr){5-7}\cmidrule(lr){8-10}
Test loading & Dataset & Exact & Shots & Kaw. & Fez & Boston & Kaw. & Fez & Boston \\
\midrule
\multirow{3}{*}{\shortstack[l]{Staircase\\(our interface)}} & Full data & 82.81 & 82.81 & 81.25 & 82.03 & 82.03 & 82.03 & 78.12 & 83.59  \\
& Random-10 & 69.53 & 71.09 & 72.66 & 69.53 & 67.19 & 71.88 & 63.28 & 71.09  \\
& QDD-10 & 83.59 & 81.25 & 77.34 & 83.59 & 82.03 & 79.69 & 78.91 & 82.03  \\
\midrule
Exact prep. & QDD-10 & 85.16 & 85.94 & 13.28 & 61.72 & 82.03 & 24.22 & 6.25 & 22.66  \\
\bottomrule
\end{tabular}
\end{center}
\end{table}

\Cref{tab:hardware} evaluates performance under exact simulation, finite-shot sampling, device noise models, and execution on three IBM quantum processors (\texttt{ibm\_kawasaki}, \texttt{ibm\_fez}, and \texttt{ibm\_boston}). We compare models trained on Full data, Random-10, and QDD-10 using identical 42-CNOT staircase loaders for the test images, so their differences reflect the training set alone. We also compare staircase loading with exact preparation of the same rank-8 targets while keeping the QDD-trained model fixed, separating the effects of training-set construction and test-time loading.

With staircase loading, the QDD-trained model remains close to the full-data model and consistently outperforms Random-10 on hardware, while all three retain much of their noiseless accuracy. The choice of test loader has a much larger effect: exact preparation achieves slightly higher noiseless accuracy, but only 6.25\% to 24.22\% on hardware, approximately 55\% to 73\% below staircase loading on the same devices and test images. The noise models substantially overestimate exact-preparation accuracy, highlighting the need for hardware validation and supporting shallow approximate loading for deploying the models.

\section{Discussion and conclusion}
\label{sec:discussion}

\textbf{Discussion.}
QDD connects training-data construction with the cost of quantum execution. Smaller training sets reduce the number of sample presentations required during training, and shallow loaders reduce the preparation cost of each shot. By optimizing samples directly within the loading-circuit family, QDD aligns the representation used during distillation with that used during training. This alignment is valuable under tight gate budgets, where post-hoc compilation can discard information from distilled states. As larger budgets allow more faithful compilation, the performance gap narrows, showing that the benefit of joint optimization depends on how strongly the circuit constrains the synthetic representation. The finite-shot comparison with random selection further demonstrates that reducing sample count must be accompanied by preserving training utility to reach high accuracy. Hardware inference extends this observation to deployment: our models remain close to full-data models through identical shallow loaders, while the loading comparison demonstrates the practical value of controlling cost. These results support jointly designing training data and loading circuits to preserve useful information within the resources available for quantum execution.

\textbf{Conclusion.}
We introduced QDD, which uses classical optimization to produce compact training sets of executable loading circuits. On MNIST and Fashion-MNIST, ten circuits per class with 42 CNOTs each provide strong predictive performance under matched sample and loading budgets. In finite-shot training on MNIST, QDD reaches 95\% of the full-data reference accuracy with more than 100-fold fewer cumulative shots than full-data full-batch training. On three IBM quantum processors, QDD-trained models remain close to full-data models under identical shallow test loaders. QDD thus provides a practical route to training data that reduces recurring quantum execution costs.

\textbf{Limitations and future work.}
Scaling distillation to larger quantum systems while retaining useful information within shallow circuits is an important open challenge. Efficient tensor-network contractions and matching selected observables offer promising directions. Additionally, applying QDD to regression and generative modeling could open new applications for compact, circuit-based training data. We hope this work inspires further research on jointly designing training data and quantum circuits for resource-efficient learning.

\bibliography{iclr2027_conference}

@article{schuld2015introduction,
  title={An introduction to quantum machine learning},
  author={Schuld, Maria and Sinayskiy, Ilya and Petruccione, Francesco},
  journal={Contemporary Physics},
  volume={56},
  number={2},
  pages={172--185},
  year={2015},
  publisher={Taylor \& Francis}
}

@article{biamonte2017quantum,
  title={Quantum machine learning},
  author={Biamonte, Jacob and Wittek, Peter and Pancotti, Nicola and Rebentrost, Patrick and Wiebe, Nathan and Lloyd, Seth},
  journal={Nature},
  volume={549},
  number={7671},
  pages={195--202},
  year={2017},
  publisher={Nature Publishing Group UK London}
}

@article{mitarai2018quantum,
  title={Quantum circuit learning},
  author={Mitarai, Kosuke and Negoro, Makoto and Kitagawa, Masahiro and Fujii, Keisuke},
  journal={Physical Review A},
  volume={98},
  number={3},
  pages={032309},
  year={2018},
  publisher={APS}
}

@article{caro2022generalization,
  title={Generalization in quantum machine learning from few training data},
  author={Caro, Matthias C and Huang, Hsin-Yuan and Cerezo, Marco and Sharma, Kunal and Sornborger, Andrew and Cincio, Lukasz and Coles, Patrick J},
  journal={Nature communications},
  volume={13},
  number={1},
  pages={4919},
  year={2022},
  publisher={Nature Publishing Group UK London}
}

@article{devadas2025quantum,
  title={Quantum machine learning: A comprehensive review of integrating AI with quantum computing for computational advancements},
  author={Devadas, Raghavendra M and Sowmya, T},
  journal={MethodsX},
  volume={14},
  pages={103318},
  year={2025},
  publisher={Elsevier}
}

@book{schuld2018supervised,
  title={Supervised learning with quantum computers},
  author={Schuld, Maria and Petruccione, Francesco},
  volume={17},
  year={2018},
  publisher={Springer}
}

@article{mottonen2004transformation,
  title={Transformation of quantum states using uniformly controlled rotations},
  author={Mottonen, Mikko and Vartiainen, Juha J and Bergholm, Ville and Salomaa, Martti M},
  journal={arXiv preprint quant-ph/0407010},
  year={2004}
}

@article{plesch2011quantum,
  title={Quantum-state preparation with universal gate decompositions},
  author={Plesch, Martin and Brukner, {\v{C}}aslav},
  journal={Physical Review A—Atomic, Molecular, and Optical Physics},
  volume={83},
  number={3},
  pages={032302},
  year={2011},
  publisher={APS}
}

@article{nakaji2022approximate,
  title={Approximate amplitude encoding in shallow parameterized quantum circuits and its application to financial market indicators},
  author={Nakaji, Kouhei and Uno, Shumpei and Suzuki, Yohichi and Raymond, Rudy and Onodera, Tamiya and Tanaka, Tomoki and Tezuka, Hiroyuki and Mitsuda, Naoki and Yamamoto, Naoki},
  journal={Physical Review Research},
  volume={4},
  number={2},
  pages={023136},
  year={2022},
  publisher={APS}
}

@article{lin2026structured,
  title={Structured Unitary Tensor Network Representations for Circuit-Efficient Quantum Data Encoding},
  author={Lin, Guang and Tanaka, Toshihisa and Zhao, Qibin},
  journal={arXiv preprint arXiv:2602.16266},
  year={2026}
}

@article{wang2018dataset,
  title={Dataset distillation},
  author={Wang, Tongzhou and Zhu, Jun-Yan and Torralba, Antonio and Efros, Alexei A},
  journal={arXiv preprint arXiv:1811.10959},
  year={2018}
}

@article{yu2023dataset,
  title={Dataset distillation: A comprehensive review},
  author={Yu, Ruonan and Liu, Songhua and Wang, Xinchao},
  journal={IEEE transactions on pattern analysis and machine intelligence},
  volume={46},
  number={1},
  pages={150--170},
  year={2023},
  publisher={IEEE}
}

@inproceedings{phalak2025dataset,
  title={Dataset distillation for quantum neural networks},
  author={Phalak, Koustubh and Li, Junde and Ghosh, Swaroop},
  booktitle={2025 IEEE Computer Society Annual Symposium on VLSI (ISVLSI)},
  volume={1},
  pages={1--5},
  year={2025},
  organization={IEEE}
}

@article{schuld2019evaluating,
  title={Evaluating analytic gradients on quantum hardware},
  author={Schuld, Maria and Bergholm, Ville and Gogolin, Christian and Izaac, Josh and Killoran, Nathan},
  journal={Physical Review A},
  volume={99},
  number={3},
  pages={032331},
  year={2019},
  publisher={APS}
}

@article{ran2020encoding,
  title={Encoding of matrix product states into quantum circuits of one-and two-qubit gates},
  author={Ran, Shi-Ju},
  journal={Physical Review A},
  volume={101},
  number={3},
  pages={032310},
  year={2020},
  publisher={APS}
}

@article{dilip2022data,
  title={Data compression for quantum machine learning},
  author={Dilip, Rohit and Liu, Yu-Jie and Smith, Adam and Pollmann, Frank},
  journal={Physical Review Research},
  volume={4},
  number={4},
  pages={043007},
  year={2022},
  publisher={APS}
}

@article{huang2024coreset,
  title={Coreset selection can accelerate quantum machine learning models with provable generalization},
  author={Huang, Yiming and Yuan, Xiao and Wang, Huiyuan and Du, Yuxuan},
  journal={Physical Review Applied},
  volume={22},
  number={1},
  pages={014074},
  year={2024},
  publisher={APS}
}

@article{oseledets2011tensor,
  title={Tensor-train decomposition},
  author={Oseledets, Ivan V},
  journal={SIAM Journal on Scientific Computing},
  volume={33},
  number={5},
  pages={2295--2317},
  year={2011},
  publisher={SIAM}
}

@article{khoromskij2011d,
  title={O (d log N)-quantics approximation of N-d tensors in high-dimensional numerical modeling},
  author={Khoromskij, Boris N},
  journal={Constructive Approximation},
  volume={34},
  number={2},
  pages={257--280},
  year={2011},
  publisher={Springer}
}

@book{nielsen2000quantum,
  title={Quantum computation and quantum information},
  author={Nielsen, Michael A and Chuang, Isaac L and others},
  volume={1},
  year={2000},
  publisher={Cambridge university press Cambridge}
}

@article{chen2012super,
  title={Super-samples from kernel herding},
  author={Chen, Yutian and Welling, Max and Smola, Alex},
  journal={arXiv preprint arXiv:1203.3472},
  year={2012}
}

@article{toneva2018empirical,
  title={An empirical study of example forgetting during deep neural network learning},
  author={Toneva, Mariya and Sordoni, Alessandro and Combes, Remi Tachet des and Trischler, Adam and Bengio, Yoshua and Gordon, Geoffrey J},
  journal={arXiv preprint arXiv:1812.05159},
  year={2018}
}

@article{sener2017active,
  title={Active learning for convolutional neural networks: A core-set approach},
  author={Sener, Ozan and Savarese, Silvio},
  journal={arXiv preprint arXiv:1708.00489},
  year={2017}
}

@article{kingma2014adam,
  title={Adam: A method for stochastic optimization},
  author={Kingma, Diederik P and Ba, Jimmy},
  journal={arXiv preprint arXiv:1412.6980},
  year={2014}
}

@article{paszke2019pytorch,
  title={Pytorch: An imperative style, high-performance deep learning library},
  author={Paszke, Adam and Gross, Sam and Massa, Francisco and Lerer, Adam and Bradbury, James and Chanan, Gregory and Killeen, Trevor and Lin, Zeming and Gimelshein, Natalia and Antiga, Luca and others},
  journal={Advances in neural information processing systems},
  volume={32},
  year={2019}
}

@article{javadi2024quantum,
  title={Quantum computing with Qiskit},
  author={Javadi-Abhari, Ali and Treinish, Matthew and Krsulich, Kevin and Wood, Christopher J and Lishman, Jake and Gacon, Julien and Martiel, Simon and Nation, Paul D and Bishop, Lev S and Cross, Andrew W and others},
  journal={arXiv preprint arXiv:2405.08810},
  year={2024}
}
\bibliographystyle{iclr2027_conference}

\appendix
\newpage

\section{State preparation and compilation}
\label{app:compilation}

QDD optimizes synthetic samples directly as staircase circuits, following the QTT representation described in \Cref{sec:background}. Row and column bits are interleaved from coarse to fine scales, with the most significant bits assigned to the highest-index qubits. This ordering organizes image structure across spatial scales along the qubit chain. The staircase depth $L$ controls representational capacity and fixes the loading cost at $2(n-1)L$ CNOTs. In the 8-qubit experiments, we use $L=3$ loaders with 42 CNOTs alongside the rank-8 input interface. Synthetic samples are learned directly through the circuit parameters, whereas real inputs and unconstrained distilled samples are converted into loaders by fitting the same circuit structure to their target states.

\textbf{Target states and compilation.}
Real inputs are normalized and truncated by TT-SVD in QTT order to the rank-8 interface. Real inputs and unconstrained distilled samples are compiled into staircase circuits by maximizing the amplitude overlap $|\langle\psi_{\mathrm{circuit}}|\psi_{\mathrm{target}}\rangle|$ with the target state. At $L=3$, the mean overlap with rank-8 test targets is 0.946 on MNIST and 0.977 on Fashion-MNIST. For unconstrained MNIST samples distilled at IPC 10, the mean overlap rises from 0.76 to 0.89, 0.96, and 0.98 as the budget grows from 14 to 28, 42, and 56 CNOTs; on Fashion-MNIST it already reaches 0.96--0.98 at 28 CNOTs across the tested IPC settings.

\section{Quantum model and resource measures}
\label{app:costmodel}

\textbf{Architecture and readout.}
The downstream model applies the block in \Cref{eq:block} on alternating even and odd nearest-neighbor bonds. With 8 qubits and 16 sublayers, it contains 56 blocks, $P=336$ trainable angles, and 112 CNOTs. Readout uses the four central qubits of the chain, which have the shortest and most symmetric light-cone distance to every input qubit under nearest-neighbor brickwork. So, measurements on qubits $\{2,3,4,5\}$ yield 16 outcomes. The first ten define the class probabilities after renormalization:
\begin{equation}
  p_\theta(c\mid\rho)
  =
  \frac{\Tr[\Pi_c W_\theta\rho W_\theta^\dagger]}
       {\sum_{c'=1}^{C}\Tr[\Pi_{c'}W_\theta\rho W_\theta^\dagger]},
  \qquad c=1,\ldots,C,
\label{eq:readout}
\end{equation}
where $\Pi_c$ projects onto the corresponding readout outcome, with the remaining qubits unobserved. Training minimizes cross-entropy with the sample labels. The 10-qubit model has 432 angles and 144 CNOTs, with readout on qubits $\{3,4,5,6\}$.

\textbf{Resource accounting.}
Each parameter-shift step evaluates $2P+1=673$ circuit settings per sample, including the unshifted circuit. A step with batch size $b$ and $S$ shots per setting therefore costs $673bS$ shots. Full-batch training uses $b=10$ at IPC 1, $b=100$ at IPC 10, and $b=60000$ for the complete dataset. The additional full-data mini-batch comparison uses $b=100$, matching the 67300 settings per step of IPC-10 training.
Shot count measures repeated circuit executions, whereas loading cost measures two-qubit gates per execution. We report loading CNOT counts before transpilation and total hardware gate counts after transpilation, including the model. At inference, each shot executes the test-image loader followed by the trained model; this cost is determined by the common test interface, independently of the training-set size.

\section{Experimental settings}
\label{app:hyper}

\textbf{Data and initialization.}
MNIST and Fashion-MNIST each contain 60000 training and 10000 test images. Images are bilinearly downsampled to $16\times16$, normalized to unit amplitude norm, and arranged in interleaved QTT order, with the coarsest row and column bits on the highest-index qubits. The 10-qubit study zero-pads the original images to $32\times32$.
QDD initializes each synthetic circuit by fitting a randomly selected real example from its assigned class. Each distillation step uses 128 examples per class and 16 randomly initialized feature circuits. The density-matrix term uses full-training-class mean states and weight $\alpha=0.3$. All optimization uses Adam \citep{kingma2014adam}.

\textbf{Baseline construction.}
All compact baselines allocate the same number of samples to each class. Random selection samples uniformly. Herding greedily matches the class mean using a linear kernel on interface states. Coreset selection retains the weighted real samples. K-center greedily selects the most distant remaining state, starting from the sample nearest the class mean. K-means runs per-class clustering on interface states and uses the renormalized centroids as samples, so it averages real images rather than selecting them.
Forgetting-based selection ranks examples by transitions from correct to incorrect classification during a seed-0 full-data training run, evaluated every 100 steps.
The distill-then-compress control optimizes unconstrained normalized vectors; when a gate budget is imposed, each optimized state is fitted to the staircase structure with the same post-hoc procedure as the selection baselines.

\textbf{Evaluation and implementation.}
Exact simulation evaluations use the complete test set and three training seeds. Finite-shot results use three training seeds and measure threshold crossings on seed-averaged curves, as detailed in \Cref{app:shot-results}.
For all experiments, we use PyTorch \citep{paszke2019pytorch} for classical optimization, and Qiskit \citep{javadi2024quantum} for quantum simulation and compilation on NVIDIA RTX A5000.

\begin{table}[t]
\caption{Finite-shot training schedules. Each training step uses $673bS$ shots, where $b$ is batch size.}
\label{tab:shot-protocol}
\renewcommand{\arraystretch}{0.9}
\begin{center}
\begin{tabular}{lrr}
\toprule
Training set & Batch size $b$ & Training steps \\
\midrule
Compact training sets, IPC 1 & 10 & 50,000 \\
Compact training sets, IPC 10 & 100 & 5,000 \\
Full data, mini-batch & 100 & 5,000 \\
Full data, full batch & 60,000 & 400 \\
\bottomrule
\end{tabular}
\end{center}
\end{table}
\begin{table}[t]
\vskip -9pt
\caption{MNIST shot savings relative to full-data full-batch training, extending \Cref{tab:budget} with full-data mini-batch training and with the best seed-averaged accuracy (Best) of every training set.}
\label{tab:budget-minibatch}
\begin{center}
\begin{tabular}{llcccc}
\toprule
Shots $S$ & Training set & 60\% & 80\% & 95\% & Best (\%) \\
\midrule
\multicolumn{6}{l}{Full-batch reference accuracy: 78.57\%} \\
\multirow{6}{*}{$100$} & Random-1 & n/a & n/a & n/a & 39.41 \\
 & Random-10 & 207$\times$ & 185$\times$ & n/a & 70.16 \\
 & QDD-1 & 1905$\times$ & n/a & n/a & 62.47 \\
 & QDD-10 & 304$\times$ & 271$\times$ & 134$\times$ & 77.19 \\
 & Full data (mini-batch) & 222$\times$ & 126$\times$ & n/a & 74.48 \\
 & Full data (full batch) & 1$\times$ & 1$\times$ & 1$\times$ & 78.57 \\
\midrule
\multicolumn{6}{l}{Full-batch reference accuracy: 81.52\%} \\
\multirow{6}{*}{$1000$} & Random-1 & n/a & n/a & n/a & 44.92 \\
 & Random-10 & 529$\times$ & 238$\times$ & n/a & 73.94 \\
 & QDD-1 & 2647$\times$ & 258$\times$ & n/a & 69.70 \\
 & QDD-10 & 529$\times$ & 380$\times$ & 576$\times$ & 81.58 \\
 & Full data (mini-batch) & 486$\times$ & 238$\times$ & 248$\times$ & 80.86 \\
 & Full data (full batch) & 1$\times$ & 1$\times$ & 1$\times$ & 81.52 \\
\midrule
\multicolumn{6}{l}{Full-batch reference accuracy: 81.02\%} \\
\multirow{6}{*}{$10000$} & Random-1 & n/a & n/a & n/a & 43.02 \\
 & Random-10 & 667$\times$ & 259$\times$ & n/a & 72.63 \\
 & QDD-1 & 3333$\times$ & 205$\times$ & n/a & 69.20 \\
 & QDD-10 & 563$\times$ & 441$\times$ & 370$\times$ & 81.44 \\
 & Full data (mini-batch) & 419$\times$ & 221$\times$ & 200$\times$ & 80.93 \\
 & Full data (full batch) & 1$\times$ & 1$\times$ & 1$\times$ & 81.02 \\
\bottomrule
\end{tabular}
\end{center}
\end{table}

\section{Additional results}
\label{app:extra}

\subsection{Finite-shot training}
\label{app:shot-results}

\textbf{Protocol.}
Models are trained from random initialization using parameter-shift gradients, and $S\in\{100,1000,10000\}$ shots per circuit setting. Compact training sets use full-batch steps. The step counts in \Cref{tab:shot-protocol} give all compact training sets the same total budget of $673\times500000\times S$ shots. Full-data mini-batch training uses the same budget and the same per-step cost as IPC-10 training.

\textbf{MNIST.}
\Cref{tab:budget-minibatch} extends the main comparison to full-data mini-batch training. With batch size fixed at 100, QDD-10 reaches the 95\% target using 2.3$\times$ and 1.9$\times$ fewer shots than mini-batch training at $S=1000$ and $10000$, respectively. At $S=100$, the mini-batch model does not reach this target. This comparison demonstrates a training-set benefit at equal cost per step; the larger savings over full-batch training also reflect the reduction from 60000 to 100 samples per step.

\subsection{Inference rank and sample count}
\label{app:rtest}

\textbf{Inference rank.}
We vary the QTT rank used to approximate test inputs while keeping the distilled sets and trained models fixed. The truncated states are evaluated directly in simulation, without additional circuit fitting. For sets distilled at rank 8, accuracy largely saturates at the same rank (\Cref{tab:interface-rank}): increasing it to 16 changes accuracy by at most 0.08\%. These results support rank 8 as an effective input approximation for the trained models.

\begin{table}[ht]
\caption{Test accuracy (\%) on MNIST versus inference rank, averaged over three training seeds.}
\label{tab:interface-rank}
\renewcommand{\arraystretch}{0.9}
\begin{center}
\begin{tabular}{lrrrrrrrr}
\toprule
IPC & $r=1$ & $r=2$ & $r=3$ & $r=4$ & $r=6$ & $r=8$ & $r=12$ & $r=16$ \\
\midrule
1 & 21.19 & 36.71 & 52.64 & 61.62 & 63.66 & 64.04 & 64.11 & 64.12 \\
10 & 25.57 & 48.52 & 64.35 & 76.84 & 81.21 & 81.62 & 81.68 & 81.68 \\
50 & 29.77 & 54.71 & 68.08 & 78.61 & 81.73 & 82.25 & 82.30 & 82.30 \\
\bottomrule
\end{tabular}
\end{center}
\end{table}
\begin{table}[ht]
\vskip -9pt
\caption{MNIST accuracy (\%) across sample budgets, averaged over three training seeds.}
\label{tab:ipc-extra}
\renewcommand{\arraystretch}{0.9}
\begin{center}
\begin{tabular}{lrrrrrr}
\toprule
Training set & IPC 1 & IPC 2 & IPC 5 & IPC 10 & IPC 20 & IPC 50 \\
\midrule
Random & 36.63 & 50.97 & 64.26 & 68.18 & 78.50 & 80.73 \\
QDD & 64.04 & 67.30 & 79.18 & 81.62 & 81.97 & 82.25 \\
\bottomrule
\end{tabular}
\end{center}
\end{table}

\textbf{Sample count.}
\Cref{tab:ipc-extra} examines the effect of the number of distilled samples per class on MNIST. QDD improves from 79.18\% at IPC 5 to 81.62\% at IPC 10, with diminishing gains thereafter: increasing the sample count fivefold to IPC 50 adds only 0.63\%. Since each additional sample incurs loading and measurement costs during full-batch training, IPC 10 offers a practical balance between predictive accuracy and training cost.

\subsection{Distillation objective}
\label{app:objectives}

{\begin{table}[htbp]
\caption{Sensitivity to the density-matrix weight on MNIST at $L=3$, IPC 10. Mean $\pm$ standard deviation over three training seeds.}
\label{tab:rho-weight}
\renewcommand{\arraystretch}{0.9}
\begin{center}
\begin{tabular}{lcccccc}
\toprule
$\alpha$ & 0.1 & 0.3 & 0.5 & 0.7 & 0.9 & 1.0 \\
\midrule
Acc. & 81.47$_{\pm0.41}$ & 81.62$_{\pm0.33}$ & 81.07$_{\pm0.70}$ & 81.31$_{\pm1.39}$ & 81.04$_{\pm0.17}$ & 81.02$_{\pm0.51}$ \\
\bottomrule
\end{tabular}
\end{center}
\end{table}
\begin{table}[!htbp]
\vskip -9pt
\caption{Effect of the density-matrix term: distribution matching alone ($\alpha=0$) versus the default objective ($\alpha=0.3$), with all other settings unchanged. Mean $\pm$ std. over three training seeds.}
\label{tab:rho-ablation}
\begin{center}
\begin{tabular}{lccc}
\toprule
Setting & DM only & DM + density-matrix & $\Delta$ \\
\midrule
IPC 1, $L=2$ & 48.15$_{\pm3.91}$ & 50.29$_{\pm1.85}$ & +2.14 \\
IPC 1, $L=3$ & 61.13$_{\pm1.44}$ & 64.04$_{\pm1.07}$ & +2.91 \\
IPC 10, $L=3$ & 81.39$_{\pm0.33}$ & 81.62$_{\pm0.33}$ & +0.23 \\
\bottomrule
\end{tabular}
\end{center}
\end{table}}

At the main setting, accuracy varies by less than 1\% across the tested density-matrix weights (\Cref{tab:rho-weight}). The $\alpha=0.3$ column coincides with the QDD entry of \Cref{tab:main}, since both use the same distilled set.
Removing the density-matrix term isolates its contribution (\Cref{tab:rho-ablation}). The term matters when each class is represented by a single circuit: at IPC 1, distribution matching alone reaches 48.15\% with $L=2$ loaders and 61.13\% with $L=3$ loaders, and anchoring the synthetic states to the density matrices raises these to 50.29\% and 64.04\%. At IPC 10 the term is neutral, changing accuracy by +0.23\% within one standard deviation, in line with the flat weight sweep above.

\subsection{10-qubit evaluation}
\label{app:n10}

We extend the evaluation to MNIST images zero-padded to $32\times32$, using 10 qubits, the same rank-8 interface, and a 16-sublayer model. Each $L=3$ loader contains 54 CNOTs. At IPC 10, QDD reaches 78.98 $\pm$ 0.87\%, which is 12.47\% above random selection and 2.95\% below the 81.93 $\pm$ 0.30\% of full-data training, as shown in \Cref{tab:n10}.

Distill-then-compress achieves 78.25 $\pm$ 0.83\% at IPC 10 and 58.27 $\pm$ 4.02\% at IPC 1 under the same objective. The similar performance of the two distilled sets is consistent with accurate post-hoc compilation at this budget.

\begin{table}[htbp]
\caption{10-qubit MNIST accuracy (\%) with 54-CNOT loaders for the compact training sets. Entries are mean $\pm$ standard deviation over three training seeds.}
\label{tab:n10}
\begin{center}
\begin{tabular}{lcc}
\toprule
Training set & IPC 1 & IPC 10 \\
\midrule
QDD & 59.12\sd{2.23} & 78.98\sd{0.87} \\
Random & 36.11\sd{1.49} & 66.51\sd{0.75} \\
Distill-then-compress & 58.27\sd{4.02} & 78.25\sd{0.83} \\
\midrule
Full data & \multicolumn{2}{c}{81.93\sd{0.30}} \\
\bottomrule
\end{tabular}
\end{center}
\end{table}

\section{Hardware evaluation and noise models}
\label{app:hardware}

\textbf{Test inputs and loading.}
Due to the high cost of running experiments on quantum hardware and to ensure a fair comparison, the models are evaluated on 128 unseen MNIST test images sampled uniformly at random. The exact-preparation control prepares the same rank-8 targets with uniformly controlled $R_y$ rotations. Every model evaluated with staircase loading receives the same fitted test circuits. These circuits encode unseen images independently of the distilled set.

\textbf{Execution and calibrations.}
We conducted evaluations of the models on both the archived model (\texttt{FakeKawasaki}, \texttt{FakeFez}, and \texttt{FakeBoston}) and real quantum hardware (\texttt{ibm\_kawasaki}, \texttt{ibm\_fez}, and \texttt{ibm\_boston}). Circuits are transpiled at optimization level 3, then execute through the Sampler with 4096 shots. Error mitigation, dynamical decoupling, and twirling are disabled.

\begin{table}[htbp]
\caption{Device error rates at hardware-job submission and in archived noise models. Two-qubit gate errors are reported as device medians.}
\label{tab:device-calibration}
\begin{center}
\begin{tabular}{lllcc}
\toprule
Device & Processor type & Role & Median 2q error & Readout error \\
\midrule
\texttt{ibm\_kawasaki} & Heron r2 & Real hardware & $1.845\times10^{-3}$ & $7.263\times10^{-3}$ \\
\texttt{ibm\_fez}      & Heron r2 & Real hardware & $2.860\times10^{-3}$ & $9.583\times10^{-3}$ \\
\texttt{ibm\_boston}   & Heron r3 & Real hardware & $1.388\times10^{-3}$ & $3.418\times10^{-3}$ \\
\midrule
\texttt{FakeKawasaki} & Eagle r3 & Archived model & $9.01\times10^{-3}$ & $15.869\times10^{-3}$ \\
\texttt{FakeFez}      & Heron r2 & Archived model & $3.90\times10^{-3}$ & $7.568\times10^{-3}$ \\
\texttt{FakeBoston}   & Heron r3 & Archived model & $1.27\times10^{-3}$ & $5.127\times10^{-3}$ \\
\bottomrule
\end{tabular}
\end{center}
\end{table}

\Cref{tab:device-calibration} summarizes the device error rates recorded at hardware-job submission and those in the archived calibrations used for noise simulation. Differences between these calibrations provide context for comparing simulated and measured performance.

\end{document}